\documentclass[twocolumn,showpacs,amsmath,amssymb,prl]{revtex4}

\usepackage{graphicx}
\usepackage{dcolumn}
\usepackage{bm}

\begin{document}

\title{Anisotropic spatially heterogeneous dynamics in a model glass-forming 
binary mixture}

\author{Elijah Flenner}
\author{Grzegorz Szamel}
\affiliation{Department of Chemistry, Colorado State University, Fort Collins, Colorado 80523}

\begin{abstract}
We calculated a four-point correlation function $G_4(\vec{k},\vec{r};t)$ and the 
corresponding structure factor $S_4(\vec{k},\vec{q};t)$ for a model glass-forming
binary mixture. These functions measure
the spatial correlations of the relaxation of different particles.
We found that these four-point functions are anisotropic and depend on the angle between 
vectors $\vec{k}$ and $\vec{r}$ (or $\vec{q}$). The anisotropy is the strongest for times somewhat longer
than the $\beta$ relaxation time but it is quite pronounced even for times comparable to the $\alpha$ relaxation
time, $\tau_{\alpha}$. At the lowest temperatures $S_4(\vec{k},\vec{q};\tau_{\alpha})$ is strongly anisotropic
even for the smallest wavevector $q$ accessible in our simulation. 
\end{abstract}

\date{\today}

\pacs{64.70.Pf, 61.20.Lc}

\maketitle 
The hypothesis that there is a growing dynamical correlation length which
accompanies the glass transition, and that this correlation length is
associated with the dramatic slowing down of a supercooled liquids'
dynamics, has recently prompted many computational and theoretical 
\cite{Bennemann1999,Donati1999,Glotzer2000,Geb2001,Lacevic2002,
Lacevic2003,Berthier2004,Toninelli2005,Doliwa2000}, 
and some experimental \cite{Berthier2006,Dauchot2005} investigations.  
Several studies have been motivated by the observation that the relaxation
of the supercooled liquid involves the correlated motion of clusters of particles,
and the size of the clusters increases with decreasing temperature
\cite{Ediger2000}.  
Since two-point correlation functions, for example the van-Hove
correlation function, do not provide any information about the 
correlated motion of particles, four-point correlation functions were introduced.  
The dynamic correlation length was determined by studying
the spatial decay of  these four-point correlation functions (or the small wavevector
dependence of their Fourier transforms), and
several studies have found that this dynamic correlation length increases 
with decreasing temperature \cite{Bennemann1999,Glotzer2000,Lacevic2002,Tracht1999}. 

The four-point correlation functions are usually assumed to be
isotropic (in some cases they are isotropic by construction \cite{Glotzer2000}).  
However, several researchers have noted that on the $\beta$ relaxation time scale
the correlated motion of particles is not isotropic.  
Doliwa and Heuer \cite{Heuer1998} observed pronounced anisotropy  
in the $\beta$ relaxation regime for a hard sphere system.
Glotzer \textit{et al.}\ reported that the motion of "mobile" particles 
for a model binary Lennard-Jones liquid is not isotropic, but the particles
move in patterns which they referred to as ``string-like'' \cite{Donati1998,Gebremichael2004}.
In other words, a particle would follow another particle in the liquid in
quasi-one-dimensional ``strings.''  These observations suggest that the
correlated motion of the liquid's molecules is anisotropic  during
$\beta$ relaxation.  Therefore, an appropriately defined four-point correlation
function used to measure this motion should also be anisotropic,
at least in the $\beta$ relaxation regime.  In this Letter we discuss 
a four-point correlation function and the corresponding structure factor which are anisotropic 
on the time scale of \emph{both} $\beta$ and $\alpha$ relaxation.  

To calculate these correlation functions we performed Brownian dynamics simulations of the Lennard-Jones
80:20 binary mixture, which was first introduced by Kob and Andersen \cite{Kob1995}.
The details of the simulations are described in 
Refs. \cite{Szamel2004,Flenner2005sim}.  In this Letter we only present results for the larger 
and more abundant A particles. Thus, all sums over particles in the formulae below
run over the A particles only.  In the figures we present the distance as $r/\sigma_{AA}$,
wavevector dependence as $q\sigma_{AA}$, and time as $t D_0/\sigma_{AA}^2$;  
$\sigma_{AA}$ is the Lennard-Jones radius for the interaction among the $A$ particles
and $D_0$ is the short time diffusion coefficient, which is temperature dependent in our simulations.

\begin{figure}
\includegraphics[scale=0.75]{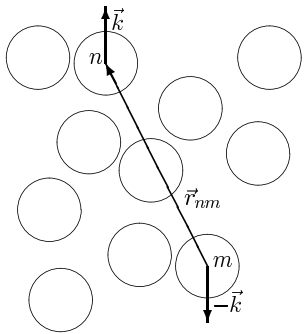}
\caption{\label{picture} $G_4(\vec{k},\vec{r};t)$ is a pair correlation function in which 
contributions of the individual particles are weighted by their microscopic self-intermediate functions 
$\hat{F}_n(\vec{k};t)$.}
\end{figure}

A four-point correlation function that we study characterizes correlations between relaxation of
different particles. Consider the function
\begin{equation}
\hat{F}_n(\vec{k};t)  =  e^{ i \vec{k} \cdot [\vec{r}_n(t) - \vec{r}_n(0)]} ,
\end{equation}
where $\vec{r}_n(t)$ is the position of particle $n$ at a time $t$.
The ensemble average of $\hat{F}_n(\vec{k};t)$
is equal to the self intermediate scattering function 
$F_s(\vec{k};t)$. The four-point function $G_4$ measures correlations between 
the microscopic self-intermediate functions pertaining to different particles separated at $t=0$
by vector $\vec{r}$ (see Fig. \ref{picture}),
\begin{equation}
G_4(\vec{k},\vec{r};t) = \frac{V}{N^2} \sum_{n \ne m} 
\left< \hat{F}_n(\vec{k};t) \hat{F}_m(-\vec{k};t) \delta[\vec{r} - \vec{r}_{nm}(0)] \right>,
\end{equation}
where $\vec{r}_{nm} = \vec{r}_n-\vec{r}_m$, $V$ is the volume, and $N$ is the number of particles.
In this work we fixed $|\vec{k}| = 7.25$,
which is around the value of the first peak in the static structure factor 
\cite{Flenner2005sim,Chandler2006}.
Note that $G_4(\vec{k},\vec{r};0) = g(r)$ where $g(r)$ is the
pair correlation function. 

The microscopic self-intermediate functions $\hat{F}_n(\vec{k};t)$ are
sensitive to particles' motions along the wavevector $\vec{k}$. Thus, the four-point 
function $G_4(\vec{k},\vec{r};t)$ measures correlations between particles' motions
along $\vec{k}$. We expect that  for small values of $|\vec{r}\,|$ the anisotropy of these correlations 
is most pronounced for $\vec{r}$ parallel to $\vec{k}$. 

To investigate local correlations on the $\alpha$ relaxation time scale \cite{comment}
we calculated projections of $G_4(\vec{k},\vec{r};t)$ onto
the Legendre polynomials,
\begin{equation}
L_n(k,r;t) = \frac{2n+1}{4\pi}
\int G_4(\vec{k},\vec{r},t) P_n(\hat{\vec{k}}\cdot\hat{\vec{r}}) d\hat{\vec{r}},
\label{leg}
\end{equation}
where $P_n$ is the $n^{th}$ Legendre polynomial, $\hat{\vec{k}}=\vec{k}/k$, 
$\hat{\vec{r}}=\vec{r}/r$, and $d\hat{\vec{r}}$ denotes integration over a unit sphere.
If $G_4(\vec{k},\vec{r};t)$ does not depend on the angle between
$\vec{k}$ and $\vec{r}$, then $L_n(k,r;t)$ will
be zero for $n > 0$.  
Note that the imaginary part of $G_4(\vec{k},\vec{r};t)$ is not 
zero, thus there are non-zero real and imaginary parts to
$L_n(k,r;t)$; by symmetry, the imaginary part of $L_n(k,r;t)$ 
is identically zero for even $n$, and the real part is zero for odd $n$. 
 
\begin{figure}
\includegraphics[scale=0.5]{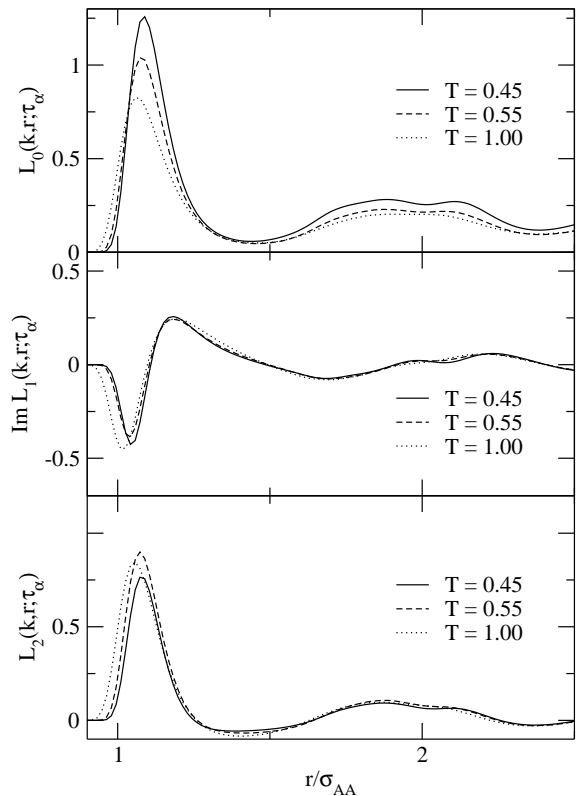}
\caption{\label{gfourp}The projections $L_0(k,r;\tau_{\alpha})$,
Im $L_1(k,r;\tau_{\alpha})$, and $L_2(k,r;\tau_{\alpha})$ for $T=0.45$, 0.55, and 1.00.}
\end{figure}

In Fig.~\ref{gfourp} we present results for $L_0(k,r;\tau_{\alpha})$, 
Im $L_1(k,r;\tau_{\alpha})$, and $L_2(k,r;\tau_{\alpha})$ 
where $\tau_\alpha$ is the $\alpha$ relaxation time \cite{comment}. As expected, 
even on this long time scale there are strong local correlations between particles' relaxation.  
The height of the first peak of the isotropic component of
$G_4$, $L_0(k,r;\tau_{\alpha})$, increases with decreasing temperature.
Furthermore, we find pronounced anisotropy of the local correlations. 
The correlations revealed by
$G_4(\vec{k},\vec{r};\tau_{\alpha})$ are the 
strongest when vectors $\vec{k}$ and $\vec{r}$ are parallel. 
The amplitude of the anisotropic part of $G_4(\vec{k},\vec{r};\tau_{\alpha})$ is, 
roughly speaking, temperature independent for $T \le 1.0$.

To examine correlations between particles' relaxation for large distances we turn to the 
structure factor corresponding to the four-point correlation function. We define 
$S_4(\vec{k},\vec{q};t) =  1 + (N/V) H_4(\vec{k},\vec{q};t)$, where $H_4(\vec{k},\vec{q};t)$ is the 
Fourier transform of  $G_4(\vec{k},\vec{r};t)-F_s^2(k;t)$. For $\vec{q}\neq \vec{0}$ we
have 
\begin{equation}
S_4(\vec{k},\vec{q};t)  = \frac{1}{N} \sum_{n, m} \left< \hat{F}_n(\vec{k};t)  \hat{F}_m(-\vec{k};t) 
e^{i \vec{q} \cdot \vec{r}_{nm}(0) } \right> .
\label{sfour} 
\end{equation}
Four-point wavevector-dependent functions similar to $S_4(\vec{k},\vec{q};t)$, 
usually evaluated at $t=\tau_{\alpha}$,
have been used previously to determine dynamic correlation lengths \cite{Bennemann1999,Lacevic2003,Toninelli2005}.

\begin{figure}
\includegraphics[scale=0.35]{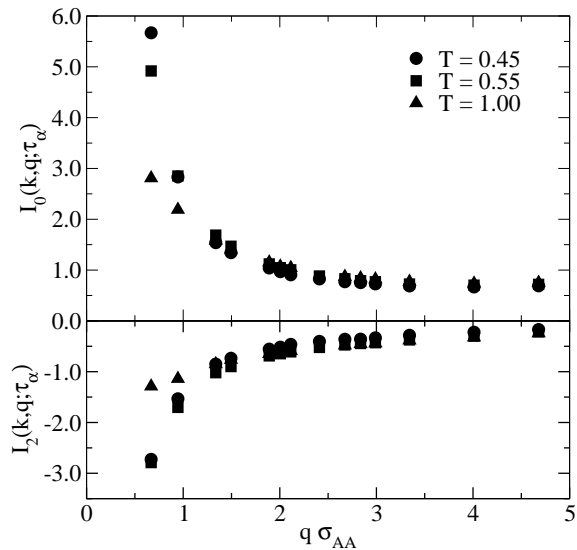}
\caption{\label{sfourp}The projections $I_0(k,q;\tau_{\alpha})$, 
and $I_2(k,q;\tau_{\alpha})$ for $T=0.45$, 0.55, and 1.00.}
\end{figure}

\begin{figure}
\includegraphics[scale=0.55]{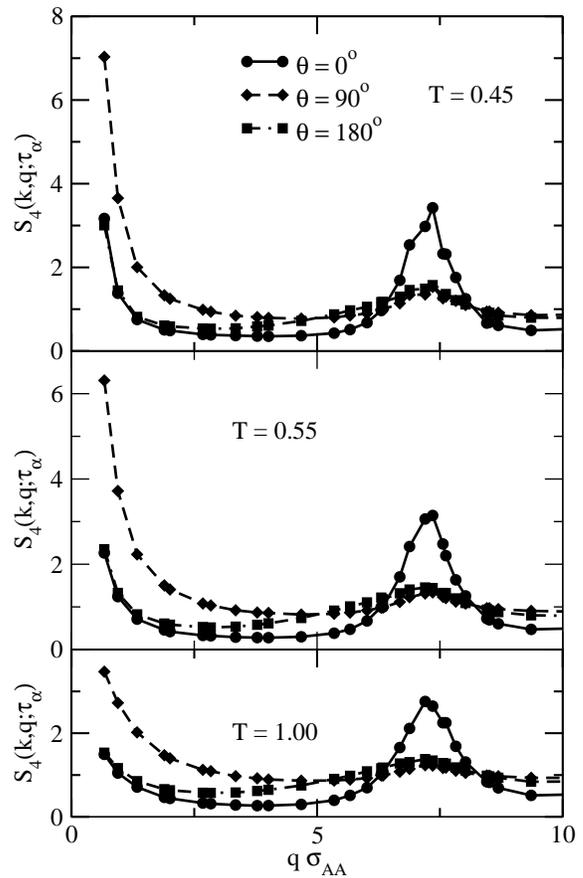}
\caption{\label{sfourq}The correlation function $S_4(\vec{k},\vec{q};\tau_\alpha)$  
for $T=0.45$, 0.55, and 1.0.
The angle $\theta$ is the angle between $\vec{k}$ and $\vec{q}$.}
\end{figure}

First, in Fig. \ref{sfourp} we examine projections of $S_4(\vec{k},\vec{q};t)$ onto
the Legendre polynomials,
\begin{equation}
I_n(k,q;t) = \frac{2n+1}{4\pi}
\int S_4(\vec{k},\vec{q},t) P_n(\hat{\vec{k}}\cdot\hat{\vec{q}}\,) d\hat{\vec{q}},
\label{sleg}
\end{equation}
at the $\alpha$ relaxation time.  We do not show $I_1(k,q;t)$ since it is small for $q\sigma_{AA}< 5$. 
Next, in Fig. \ref{sfourq} we examine $S_4(\vec{k},\vec{q};\tau_{\alpha})$ for different angles
$\theta$ between $\vec{k}$ and $\vec{q}$. In accordance with numerous earlier
investigations \cite{Bennemann1999,Donati1999,Geb2001,Lacevic2002},
 we find that the large
long-range correlations between particles' relaxation increase with decreasing
temperature. Surprisingly, we find that even the long-range correlations on the $\alpha$ relaxation
time scale are strongly anisotropic. At low $q$ the correlations are strongest for $\vec{q}\perp\vec{k}$.
In direct space this corresponds to strong correlations for particles separated by a vector $\vec{r}$ which is parallel to $\vec{k}$.

The anisotropy of $S_4(\vec{k},\vec{q};t)$ makes the determination of a single dynamic
correlation length difficult. We tried to fit the low
$\vec{q}$ values of $S_4(\vec{k},\vec{q};t)$ for a fixed angle 
$\theta$ between $\vec{k}$ and $\vec{q}$ to several different functional forms, 
but could not find a form where the results were reasonable for $T < 0.50$.
Specifically, fits to an Ornstein-Zernicke formula 
were very poor. The correlation length determined from the fits was either larger than 
half the simulation cell, or we had to fix the unknown $\vec{q} = \vec{0}$ value of 
$S_4(\vec{k},\vec{q};t)$.
Larger systems need to be simulated in order to better understand the low $q$ behavior
 of $S_4(\vec{k},\vec{q};t)$.

\begin{figure}
\includegraphics[scale=0.25]{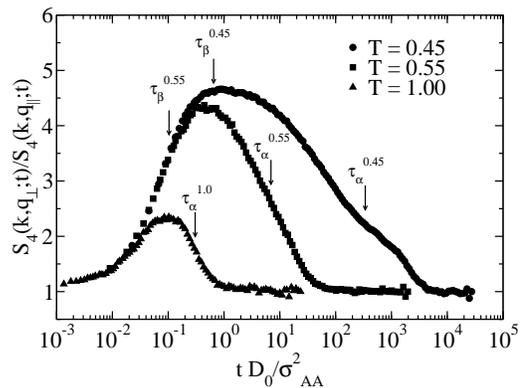}
\caption{\label{anisotropy}The ratio $S_4(\vec{k},q_{\perp};t)/S_4(\vec{k},q_{\parallel};t)$ 
for $T=1.0$, 0.55, and 0.45.  The arrows indicate
the $\alpha$ and the $\beta$ relaxation times.}
\end{figure}

While it has been previously shown that heterogeneous dynamics on the $\beta$ relaxation
time scale is anisotropic \cite{Heuer1998}, the strong anisotropy of $S_4(\vec{k},\vec{q};\tau_{\alpha})$
was a surprise. To investigate the time dependence of the anisotropy we examined the ratio
$S_4(\vec{k},\vec{q}_{\perp};t)/S_4(\vec{k},\vec{q}_{\parallel};t)$, where $S_4(\vec{k},\vec{q}_{\perp};t)$
is calculated for the smallest wavevector allowed due to the periodic boundary 
conditions ($|\vec{q}_{\perp}| = 2\pi/L$) perpendicular to $\vec{k}$ and $S_4(\vec{k},\vec{q}_{\parallel};t)$
is calculated for the smallest wavevector ($|\vec{q}_{\parallel}| = 2\pi/L$) parallel to $\vec{k}$.  
This ratio is shown in 
Fig.~\ref{anisotropy} for $T=1.0$, 0.55, and 0.45.  The arrows in the figure
indicate the $\alpha$ relaxation time and the $\beta$ relaxation time, $\tau_\beta$.
The $\beta$ relaxation time was determined by finding the first inflection
point of $\ln[F_s(\vec{k};t)]$ as a function of $t$.  For $T=1.0$, this inflection point does not 
exist.  The peak in the ratio $S_4(\vec{k},\vec{q}_{\perp};t)/S_4(\vec{k},\vec{q}_{\parallel};t)$
(\textit{i.e.} the maximum anisotropy) occurs between the $\beta$ and the $\alpha$ 
relaxation time, but is closer to the $\beta$ relaxation time for the lower temperatures. However,
the correlated motion is still strongly anisotropic around $\tau_\alpha$. 

\begin{figure}
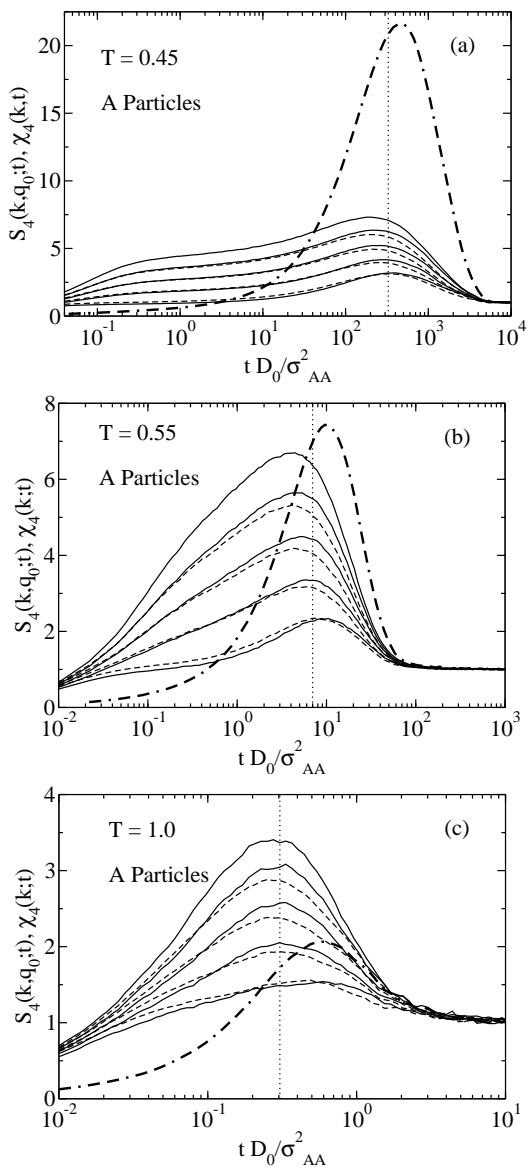

\includegraphics[scale=0.25]{fpangle045n3.eps}\\[0.1cm]
\includegraphics[scale=0.25]{fpangle055n3.eps}\\[0.1cm]
\includegraphics[scale=0.25]{fpangle10n3.eps}
\caption{\label{sfourfig}The four-point correlation function $S_4(\vec{k},\vec{q}_0;t)$
as a function of time for
T=0.45, 0.55, and 1.0.  The solid lines are for the
angles $0^\circ$, $30^\circ$, $45^\circ$, 
$60^\circ$, and $90^\circ$ listed from bottom to top.  
The dashed lines represent the angles $120^\circ$, $135^\circ$, 
$150^\circ$, and $180^\circ$ listed from top to bottom.  The dark dashed dotted line is $\chi_4(k;t)$
and the vertical dotted line marks the $\alpha$
relaxation time.}
\end{figure}
 
Finally, we compared $S_4(\vec{k},\vec{q}_0;t)$ for the smallest
wavevector allowed due to the periodic boundary conditions, $|\vec{q}_0| = 2\pi/L$, with 
the four-point susceptibility
\begin{eqnarray}
\chi_4(k;t) & = & \frac{1}{N} \sum_{n, m} \left< \hat{F}_n(\vec{k};t) \hat{F}_m(-\vec{k};t) \right> \nonumber \\
& & - \left<\hat{F}_n(\vec{k};t) \right> \left< \hat{F}_m(-\vec{k};t) \right>.
\label{sus}
\end{eqnarray}
As observed before \cite{Chandler2006}, while $S_4(\vec{k},\vec{q}=0;t)=\chi_4(k;t)$, the susceptibility 
$\chi_4(k;t)$ is ensemble-dependent, and in a constant $N$ ensemble 
$\lim_{\vec{q}\rightarrow 0}S_4(\vec{k},\vec{q};t)\neq \chi_4(k;t)$. In Fig.~\ref{sfourfig}
we show the correlation function $S_4(\vec{k},\vec{q}_0;t)$ and the susceptibility 
$\chi_4(k;t)$ for several angles $\theta$ 
between vectors $\vec{q}_0$ and $\vec{k}$ for $T=1.0$, 0.55, and 0.45.
For a  fixed $\theta$ there is a peak in $S_4(\vec{k},\vec{q}_0;t)$ 
around $\tau_{\alpha}$. The exact position of the peak
depends on the angle between $\vec{q}_0$ and $\vec{k}$,
and occurs at a time larger than 
$\tau_\alpha$ for $\vec{q}_0$ parallel to $\vec{k}$, but smaller
than the $\tau_\alpha$ for $\vec{q}_0$ perpendicular to $\vec{k}$. 
It is clear from Fig. \ref{sfourfig} that the time dependence of the correlation function 
$S_4(\vec{k},\vec{q}_0;t)$ and that of the susceptibility $\chi_4(k;t)$
are quite different. This is significant because time dependence predicted by recent
extensions of the mode-coupling theory \cite{BBEPL} for the correlation
function $S_4(\vec{k},\vec{q};t)$ are usually checked against simulation results for
the susceptibility $\chi_4(k;t)$ \cite{Toninelli2005,Szamel2006}.

In summary, we have shown that there is a pronounced anisotropy of the correlations of particles' relaxation
in a model supercooled liquid even on the time scale of the $\alpha$ relaxation time. 
This anisotropy will need to be addressed
in theoretical descriptions of heterogeneous dynamics of supercooled liquids. In particular
theoretical input is needed in order to elucidate the relation between anisotropic 
four-point correlation functions and a dynamic correlation length. Finally, 
we expect that the anisotropy of the four-point correlation functions examined in this Letter
will help to differentiate between competing theoretical descriptions of heterogeneous dynamics.   

We gratefully acknowledge the support of NSF Grant No.~CHE 0517709.


\begin{thebibliography}{99}
\bibitem{Bennemann1999}C. Bennemann, C. Donati, J. Baschnagel, and S.C. Glotzer, 
  Nature \textbf{399}, 246 (1999).
\bibitem{Donati1999}C. Donati, S.C. Glotzer, P.H. Poole, W. Kob, and S.J. Plimpton, 
  Phys. Rev. E \textbf{60}, 3107 (1999).
\bibitem{Glotzer2000} S.C. Glotzer, V.N. Novikov, and T.B. Schroder, 
  J. Chem. Phys. \textbf{112}, 509 (2000).
\bibitem{Doliwa2000} B. Doliwa, and A. Heuer, 
  Phys. Rev. E \textbf{61}, 6898 (2000).
\bibitem{Geb2001}Y. Gebremichael, T.B. Schroder, F.W. Starr, and S.C. Glotzer, 
  Phys. Rev. E \textbf{64}, 051503 (2001).
\bibitem{Lacevic2002}N. Lacevic, F.W. Starr, T.B. Schroder, V.N. Novikov, and S.C. Glotzer, 
  Phys. Rev. E \textbf{66}, 030101 (2002).
\bibitem{Lacevic2003} N. Lacevic, F.W. Starr, T.B. Schroder, and S.C. Glotzer, 
  J. Chem. Phys. \textbf{119}, 7372 (2003).
\bibitem{Berthier2004} L. Berthier, 
  Phys. Rev. E \textbf{69}, 020201(R) (2004).
\bibitem{Toninelli2005} C. Toninelli, M. Wyart, L. Berthier, G. Biroli, and J.P. Bouchaud, 
  Phys. Rev. E \textbf{71}, 041505 (2005).\bibitem{Berthier2006} L. Berthier, G. Biroli, J.P. Bouchaud, L. Cipelletti, D. El Masri, D. Lhote, F. Ladieu, M. Pierno, 
 Science \textbf{310}, 1797 (2006).
\bibitem{Dauchot2005} O. Dauchot, G. Marty, and G. Biroli, 
 Phys. Rev. Lett. \textbf{92}, 185705 (2005).
\bibitem{Ediger2000} M.D. Ediger, 
 Annu. Rev. Phys. Chem. \textbf{51}, 99 (2000).
\bibitem{Tracht1999} U. Tracht, M. Wilhelm, A. Heuer, H.W. Speiss, 
 J. Magn. Reson. \textbf{140}, 460 (1999).
\bibitem{Heuer1998} B. Doliwa, and A. Heuer, 
 Phys. Rev. Lett. \textbf{80}, 4915 (1998).
\bibitem{Donati1998} C. Donati, J.F. Douglas, W. Kob, S.J. Plimpton, P.H. Poole, and S.C. Glotzer,
 Phys. Rev. Lett. \textbf{80}, 2338 (1998).
\bibitem{Gebremichael2004} Y. Gebremichael, M. Vogel, and S.C. Glotzer,
 J. Chem. Phys. \textbf{120}, 4415 (2004).
\bibitem{Kob1995} W. Kob, and H.C. Andersen, 
 Phys. Rev. E \textbf{51}, 4626 (1995); \textbf{52} 4134(1995)
\bibitem{Szamel2004} G. Szamel, and E. Flenner,
  Europhys. Lett. \textbf{67}, 779 (2004).
\bibitem{Flenner2005sim} E. Flenner and G. Szamel,
 Phys. Rev. E \textbf{72}, 031508 (2005); Phys. Rev. E \textbf{72}, 011205 (2005).
\bibitem{Chandler2006}The susceptibility $\chi_4(k;t)$ depends
 non-trivially on $k$, see D. Chandler, J.P. Garrahan, R.L. Jack, L. Maibaum, and A.C. Pan, 
 cond-mat/0605084 (2006).
\bibitem{comment}The $\alpha$ relaxation time is defined to be when 
 $F_s(\vec{k};\tau_\alpha) = e^{-1}$, where $|\vec{k}| = 7.25$.
 \bibitem{Biroli2006} G. Biroli, J.P. Bouchaud, K. Miyazaki, and D.R. Reichman, 
 cond-mat/0605733 (2006).
\bibitem{BBEPL} G. Biroli and J.-P. Bouchaud, Europhys. Lett. \textbf{67},
21 (2004).
 \bibitem{Szamel2006} G. Szamel and E. Flenner, Phys. Rev. E \textbf{74}, 021507 (2006).
 \end{thebibliography}
\end{document}